\documentclass[11pt,twoside]{article}
\usepackage{ap-article}
\def\labart{yourLabel}      % put a label from your choice here
\shortauthors{M. Borg \textit{et al.}}
\shorttitle{Safely Entering the Deep}
\title{Safely Entering the Deep: A Review of Verification and Validation for Machine Learning and a Challenge Elicitation in the Automotive Industry}
\author{%
Markus Borg\,\up{1}\,
Cristofer Englund\,\up{1},
Krzysztof Wnuk\,\up{2},
Boris Duran\,\up{1},
Christoffer Levandowski\,\up{3},   
Shenjian Gao\,\up{2},
Yanwen Tan\,\up{2},
Henrik Kaijser\,\up{4},
Henrik L\"onn\,\up{4},
Jonas T\"ornqvist\,\up{3}}
\addresses{%
\up{1}
RISE Research Institutes of Sweden AB,\\
Scheelev\"agen 17,\\
SE-223 70 Lund, Sweden
\\ \vspace*{0.04truein}
E-mail: \{markus.borg, cristofer.englund, boris.duran\}@ri.se
\\ \vspace*{0.05truein}
\up{2}
Blekinge Institute of Technology,\\
Valhallav\"agen 1,\\ SE-371 41 Karlskrona, Sweden \\ 
\vspace*{0.04truein}
E-mail: krzysztof.wnuk@bth.se, \{shga13, yata13\}@student.bth.se
\\ \vspace*{0.05truein}
\up{3}
QRTECH AB,\\
Fl\"ojelbergsgatan 1C,\\ SE-431 35 M\"olndal, Sweden \\ 
\vspace*{0.04truein}
E-mail: \{christoffer.levandowski, jonas.tornqvist\}@qrtech.se
\\ \vspace*{0.05truein}
\up{4}
AB Volvo,\\
Volvo Group Trucks Technology,\\ SE-405 08 Gothenburg, Sweden \\ 
\vspace*{0.04truein}
E-mail: \{henrik.kaijser, henrik.lonn\}@volvo.se
}
\begin{document}
\label{\labart-FirstPage}

\maketitle
%-------------------------- ABSTRACT ---------------------------------------
\begin{abstract} %MAX 80 WORDS IN JASE
Deep Neural Networks (DNN) will emerge as a cornerstone in automotive software engineering. However, developing systems with DNNs introduces novel challenges for safety assessments. This paper reviews the state-of-the-art in verification and validation of safety-critical systems that rely on machine learning. Furthermore, we report from a workshop series on DNNs for perception with automotive experts in Sweden, confirming that ISO~26262 largely contravenes the nature of DNNs. We recommend aerospace-to-automotive knowledge transfer and systems-based safety approaches, e.g., safety cage architectures and simulated system test cases.
\end{abstract}

\medskip
%-------------------------- KEYWORDS ---------------------------------------

\vspace*{7pt}\textlineskip
%-------------------------- BEGIN BODY OF TEXT -----------------------------
\begin{multicols}{2}

\section{Introduction}
As an enabling technology for autonomous driving, Deep learning Neural Networks (DNN) will emerge as a cornerstone in automotive software engineering. Automotive software solutions using DNNs is a hot topic, with new advances being reported almost weekly. Also in the academic context, several research communities study DNNs in the automotive domain from various perspectives, e.g., applied Machine Learning (ML)\cite{ramos_detecting_2016}, software engineering\cite{falcini_deep_2017}, safety engineering\cite{salay_analysis_2017}, and Verification \& Validation (V\&V)\cite{katz_reluplex:_2017}.

DNNs are used to enable \textit{vehicle environmental perception}, i.e., awareness of elements in the surrounding traffic. Successful perception is a prerequisite for autonomous features such as lane departure detection, path/trajectory planning, vehicle tracking, behavior analysis, and scene understanding\cite{zhu_overview_2017} -- and a prerequisite to reach levels 3-5 as defined by SAE International's levels of driving automation. A wide range of sensors have been used to collect input data from the environment, but the most common approach is to rely on front-facing cameras\cite{gurghian_deeplanes:_2016}. In recent years, DNNs have demonstrated their usefulness in classifying such camera data, which in turn has enabled both perception and subsequent breakthroughs toward autonomous driving\cite{lecun_deep_2015}.

From an ISO~26262 safety assurance perspective, however, developing systems based on DNNs constitutes a major paradigm shift compared to conventional systems\footnote{by \textit{conventional systems} we mean any system that does not have the ability to learn or improve from experience}~\cite{falcini_deep_2017}. Andrej Karpathy, Director of AI at Tesla, boldly refers to the new era as ``Software 2.0''\footnote{https://medium.com/@karpathy/software-2-0-a64152b37c35}. No longer do human engineers explicitly describe all system behavior in source code, instead DNNs are trained using enormous amounts of historical data. 

DNNs have been reported to deliver superhuman classification accuracy for specific tasks\cite{he_delving_2015}, but inevitably they will occasionally fail to generalize\cite{spanfelner_challenges_2012}. Unfortunately, from a safety perspective, analyzing when this might happen is currently not possible due to the black-box nature of DNNs. A state-of-the-art DNN might be composed of hundreds of millions of parameter weights, thus the methods for V\&V of DNN components must be different compared to approaches for human readable source code. Techniques enforced by ISO~26262 such as source code reviews and exhaustive coverage testing are not applicable\cite{salay_analysis_2017}. 

The contribution of this review paper is twofold. First, we describe the state-of-the-art in V\&V of safety-critical systems that rely on ML. We survey  academic literature, partly through a reproducible snowballing  review~\cite{wohlin_guidelines_2014}, i.e., establishing a body of literature by tracing referencing and referenced papers. Second, we elicit the most pressing challenges when engineering safety-critical DNN components in the automotive domain. We report from workshops with automotive experts, and we validate findings from the literature review through an industrial survey. The research  has been conducted as part of SMILE\footnote{The SMILE project: Safety analysis and verification/validation of MachIne LEarning based systems}, a joint research project between RISE AB, Volvo AB, Volvo Cars, QRTech AB, and Semcon AB. 

The rest of the paper is organized as follows: Section~\ref{sec:background} presents safety engineering concepts within the automotive domain and introduces the fundamentals of DNNs. Section~\ref{sec:method} describes the proposed research method, including four sources of empirical evidence, and Section~\ref{sec:res} reports our findings. Section~\ref{sec:synth} presents a synthesis targeting our two objectives, and discusses implications for research and practice. Finally, Section~\ref{sec:concl} concludes the paper and outlines the most promising directions for future work. Throughout the paper, we use the notation \textbf{[PX]} to explicitly indicate publications that are part of the snowballing literature study.

\section{Background}\label{sec:background}
This section first presents development of safety-critical software according to the ISO~26262 standard\cite{international_organization_for_standardization_iso_2011}. Second, we introduce fundamentals of DNNs, required to understand how it could allow vehicular perception. In the remainder of this paper, we adhere to the following three definitions related to safety-critical systems:
\begin{itemize}
\item \textbf{Safety} is ``freedom from unacceptable risk of physical injury or of damage to the health of people''\cite{international_electrotechnical_commission_iec_2010}
\item \textbf{Robustness} is ``the degree to which a component can function correctly in the presence of invalid inputs or stressful environmental conditions''\cite{ieee_computer_society_610.12-1990_1990}
\item \textbf{Reliability} is ``the probability that a component performs its required functions for a desired period of time without failure in specified environments with a desired confidence''\cite{billinton_reliability_2013}
\end{itemize}

\subsection{Safety Engineering in the Automotive Domain: ISO~26262}
\label{sec:safety}
Safety is not a property that can be added at the end of the design. Instead, it must be an integral part of the entire engineering process. To successfully engineer a safe system, a systematic safety analysis and a methodological approach to managing risks are required\cite{bahr_system_2014}. Safety analysis comprises identification of hazards, development of approaches to eliminate hazards or mitigate their consequences, and verification that the approaches are in place in the system. Risk assessment is used to determine how safe a system is, and to analyze alternatives to lower the risks in the system.

Safety has always been an important concern in engineering, and best practices have often been collected in governmental or industry \textit{safety standards}. Common standards provide a common vocabulary as well as a way for both internal and external safety assessment, i.e., work tasks for both engineers working in the development organization and for independent safety assessors from certification bodies. For software-intensive systems, the generic meta-standard IEC~61508\cite{international_electrotechnical_commission_iec_2010} introduces the fundamentals of functional safety for Electrical/Electronic/Programmable Electronic (E/E/PE) Safety-related Systems, i.e., hazards caused by malfunctioning E/E/PE systems rather than non-functional considerations such as fire, radiation, and corrosion. Several different domains have their own adaptations of IEC~61508.

ISO~26262\cite{international_organization_for_standardization_iso_2011} is the automotive derivative of IEC~61508, organized into 10 parts, constituting a comprehensive safety standard covering all aspects of automotive development, production, and maintenance of safety-related systems. V\&V are core activities in safety-critical development and thus discussed in detail in ISO~26262, especially in Part 4: Product development at the system level and Part 6: Product development at the software level. The scope of the current ISO~26262 standard is series production passenger cars with a max gross weight of 3,500~kg. However, the second edition of the standard, expected in the beginning of 2019, will broaden the scope to cover also trucks, buses, and motorcycles.

The \textit{automotive safety lifecycle} (ASL) is one key component of ISO~26262\cite{national_instruments_what_nodate}, defining fundamental concepts such as safety manager, safety plan, and confirmation measures including safety review and audit. The ASL describes six phases: management, development, production, operation, service, and decommission. Assuming that a safety-critical DNN will be considered a software unit, especially the development phase on the software level (Part 6) mandates practices that will require special treatment. Examples include verification of software implementation using inspections (Part 6:8.4.5) and conventional structural code coverage metrics (Part 6:9.4.5). It is evident that certain ISO~26262 process requirements cannot apply to ML-based software units, in line with how model-based development is currently partially excluded.

Another key component of ISO~26262 is the \textit{automotive safety integrity level} (ASIL). In the beginning of the ASL development phase, a safety analysis of all critical functions of the system is conducted, with a focus on hazards. Then a risk analysis combining 1) the probability of exposure, 2) the driver's possible controllability, and 3) the possible severity of the outcome, results in an ASIL between A and D. ISO~26262 enforces development and verification practices corresponding to the ASIL, with the most rigorous practices required for ASIL D. Functions that are not safety-critical, i.e., below ASIL A, are referred to as `QM' as no more than the normal quality management process is enforced.

\subsection{Deep Learning for Perception: Approaches and Challenges}
\label{sec:dnn}
While there currently is a deep learning hype, there is no doubt that the technique has produced groundbreaking results in various fields -- by clever utilization of the increased processing power in the last decade, nowadays available in inexpensive GPUs, combined with the ever-increasing availability of data.

Deep learning is enabled by DNNs, which are a kind of Artificial Neural Networks (ANN). To some extent inspired by biological connectomes, i.e., mappings of neural connections such as in the human brain, ANNs composed of connected layers of neurons are designed to learn to perform classification tasks. While ANNs have been studied for decades, significant breakthroughs came when the increased processing power allowed adding more and more layers of neurons -- which also increased the number of connections between neurons by orders of magnitude. The exact number of layers, however, needed for a DNN to qualify as deep is debatable.  

A major advantage of DNNs is that the classifier is less dependent on \textit{feature engineering}, i.e., using domain knowledge to (perhaps manually) identify properties in data for ML to learn from -- this is often difficult. Examples of operations used to extract features in computer vision include: color analysis, edge extraction, shape matching, and texture analysis. What DNNs instead introduced was an ML solution that learned those features directly from input data, greatly decreasing the need for human feature engineering. DNNs have been particularly successful in speech recognition, computer vision, and text processing -- areas in which ML results were limited by the tedious work required to extract effective features.

In computer vision, essential for vehicular perception, the state-of-the-art is represented by a special class of DNNs known as \textit{Convolutional Neural Networks} (CNN) \cite{he_deep_2016,szegedy_going_2015,simonyan_very_2014,donahue_decaf:_2014}. Since 2010, several approaches based on CNNs have been proposed -- and in only five years of incremental research the best CNNs matched the image classification accuracy of humans. CNN-based image recognition is now reaching the masses, as companies like Nvidia, Intel, etc. are now commercializing specialized hardware with automotive applications in mind such as the Drive PX series. Success stories in the automotive domain include lane keeping applications for self-driving cars\cite{bojarski_end_2016,sallab_end--end_2016}.

\textit{Generative Adversarial Networks} (GAN) is another approach in deep learning research that is currently receiving considerable interest\cite{goodfellow_nips_2016,radford_unsupervised_2015}. In contrast to discriminative networks (what has been discussed so far) that learn boundaries between classes in the data for the purpose of classification, a generative network can instead be used to learn the probability of features given a specific class. Thus, a GAN could be used to generate samples from a learned network -- which could possibly be used to expand available training data with additional synthetic data. GANs can also be used to generate \textit{adversarial examples}, i.e., inputs to ML classifiers intentionally created to cause misclassification.  

Finally, successful applications of DNNs rely on the availability of large labeled datasets from which to learn features. In many cases, such labels are limited or does not exist at all. To maximize the utility of the labeled data, truly hard currency for anyone engineering ML-based systems, techniques such as \textit{transfer learning} are used to adapt knowledge learned from one dataset to another domain\cite{glorot_domain_2011}.

\section{Research method}\label{sec:method}
The overarching goal of the SMILE project is to develop approaches to V\&V of ML-based systems, more specifically automotive applications relying on DNNs. Our current paper is guided by two research questions:

\begin{itemize}
\item[RQ1] What is the state-of-the-art in V\&V of ML-based safety-critical systems?
\item[RQ2] What are the main challenges when engineering safety-critical systems with DNN components in the automotive domain?
\end{itemize}

Fig.~\ref{fig:processOverview} shows an overview of the research, divided into three sequential parts (P1-P3). Each part concluded with a Milestone (I--III).
In Fig.~\ref{fig:processOverview}, tasks driven by academia (or research institutes) are presented in the light gray area -- primarily addressing RQ1. Tasks in the darker gray area above, are primarily geared toward collecting data in the light of RQ2, and mostly involve industry practitioners. The darkest gray areas denote involvement of practitioners that were active in safety-critical development but not part of the SMILE project.

\begin{Figure}
\centering
\includegraphics[width=2.2\textwidth]{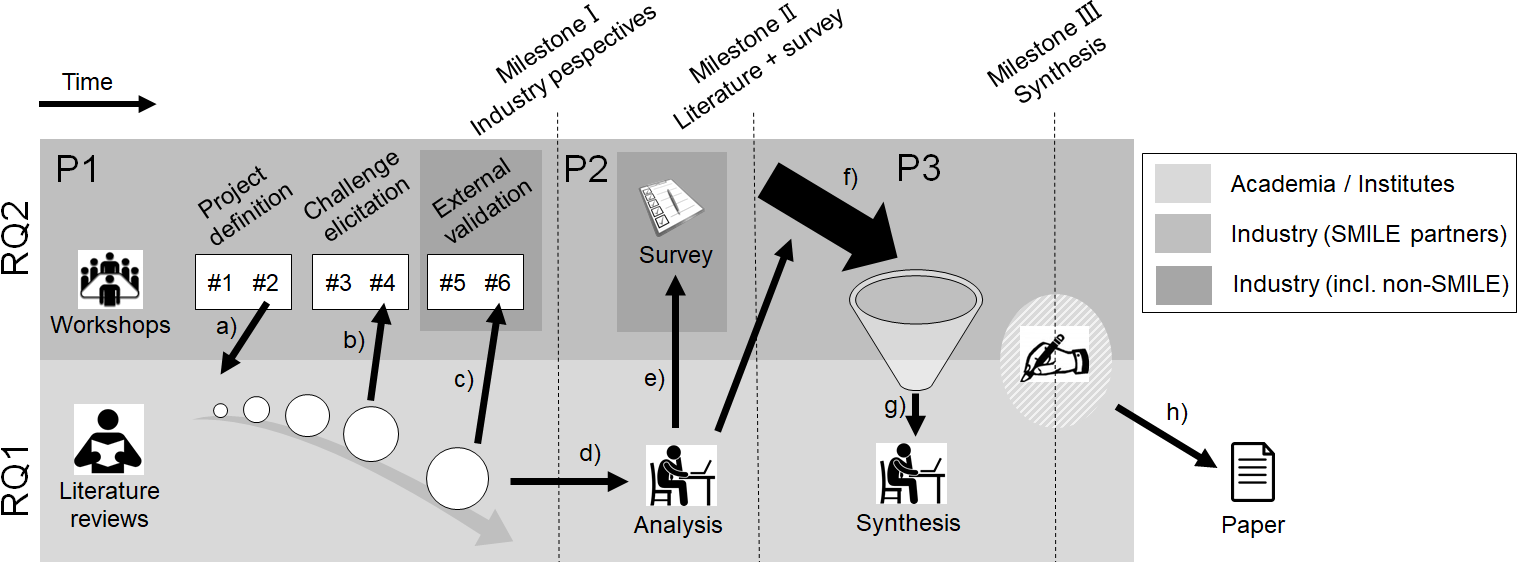}
\fcaption{Overview of the SMILE project and its three milestones. The figure illustrates the joint industry/academia nature of SMILE, indicated by light gray background for tasks driven by academia and darker gray for tasks conducted by practitioners.}
%The darkest gray background show tasks involving practitioners outside of the project, strengthening the external validity of our findings.}
\label{fig:processOverview}
\end{Figure}
%\vspace*{2cm}

In the first part of the project (P1 in Fig.~\ref{fig:processOverview}), we initiated a systematic snowballing review of academic literature to map the state-of-the-art. In parallel, we organized a workshop series with domain experts from industry with monthly meetings to also assess the state-of-practice in the Swedish automotive industry. The literature review was seeded by discussions from the project definition phase (a). Later, we shared intermediate findings from the literature review at workshop \#4 (b) and final results were brought up to discussion at workshop \#6 (c). The first part of the project concluded with Milestone~I: a collection of industry perspectives.

The second part of the SMILE project (P2 in Fig.~\ref{fig:processOverview}) involved an analysis of the identified literature (d). We extracted challenges and solution proposals from the literature, and categorized them according to a structure that inductively emerged during the process (see Section~\ref{sec:slr}). Subsequently, we created a questionnaire-based survey to validate our findings and to receive input from industry practitioners beyond SMILE (e). The second phase concluded with analyzing the survey data at Milestone~II.%, resulting in a MSc thesis project report [REF] (not represented in Fig.~\ref{fig:processOverview}). \todo[inline]{If it is NOT in the figure why even mentioning it?}

\newpage
In the third part of the project (P3 in Fig.~\ref{fig:processOverview}), we collected all results (f), and performed a synthesis (g). Finally, writing this article concludes the research at Milestone~III.

Fig.~\ref{fig:evidence} shows an overview of the SMILE project from an evidence perspective. The collection of empirical evidence was divided into two independent tracks resulting in four sets of evidence, reflecting the nature of the joint academia/industry project. Furthermore, the split enabled us to balance the trade-off between rigor and relevance that plagues applied research projects\cite{ivarsson_method_2011}.

As shown in the upper part of Figure~\ref{fig:evidence}, the SMILE consortium performed (non-replicable, from now on: ``\textit{ad hoc}'') searching for related work. An early set of papers was used to seed the systematic search described in the next paragraph. The findings in the body of related work (cf. A in Fig.~\ref{fig:evidence}) were discussed at the workshops. The workshops served dual purposes, they collected empirical evidence of priorities and current needs in the Swedish automotive industry (cf. B in Fig.~\ref{fig:evidence}), and they validated the relevance of the research identified through the \textit{ad hoc} literature search. %Note that additional pieces of related work suggested by the reviewers were added to (cf. h) in Fig.~\ref{fig:processOverview}).
The upper part focused on \textit{maximizing industrial relevance}, at the expense of rigor, i.e., we are certain that the findings are relevant to the Swedish automotive industry, but the research was conducted in an \textit{ad hoc} fashion with limited traceability and replicability. The right part of Figure~\ref{fig:evidence} complements the practice-oriented research of the SMILE project by a systematic literature review, adhering to an established process\cite{wohlin_guidelines_2014}. The identified papers (cf. C in Fig.~\ref{fig:evidence})) were systematized and the result was validated through a questionnaire-based survey. The survey also acted as a means to collect additional primary evidence, as we collected practitioners' opinions on V\&V of ML-based systems in safety-critical domains (cf. D in Fig.~\ref{fig:evidence}). Thus, the lower part focused on \textit{maximizing academic rigor}. %Note that the right part was not limited to the automotive domain -- instead we increased the scope to cover ML in any safety-critical application.

\begin{Figure}
\centering
\includegraphics[width=0.99\textwidth]{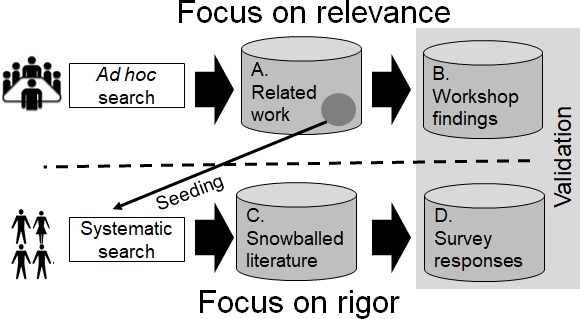}
\fcaption{Overview of the SMILE project from an evidence perspective. We treat the evidence as four different sets: A. Related work and C. Snowballed literature represent secondary evidence, whereas B. Workshop findings and D. Survey responses constitute primary evidence.}
\label{fig:evidence}
\end{Figure}

\subsection{The systematic review} \label{sec:slr}
Inspired by evidence-based medicine, systematic literature reviews have become a popular software engineering research method to aggregate work in a research area. Snowballing literature reviews\cite{wohlin_guidelines_2014} is an alternative to more traditional database searches relying on carefully developed search strings, particularly suitable when the terminology used in the area is diverse, e.g., in early stages of new research topics. This section describes the two main phases of the literature review: 1) paper selection and 2) data extraction and analysis.

\subsubsection{Paper selection}
As safety-critical applications of DNNs in the automotive sector is still a new research topic, we decided to broaden our literature review to encompass also other types of ML, and also to go beyond the automotive sector. We developed the following criteria: for a publication to be included in our literature review, it should describe 1) engineering of an ML-based system 2) in the context of autonomous cyber-physical systems, and 3) the paper should address V\&V or safety analysis. Consequently, our criteria includes ML beyond neural networks and DNNs.
Our focus on autonomous cyber-physical systems implicitly restricts our scope to safety-critical systems. Finally, we exclude papers that do not target V\&V or safety analysis, but instead other engineering considerations, e.g., requirements engineering, software architecture, or implementation issues. 

First, we established a \textit{start set} using exploratory searching in Google Scholar and applying our inclusion criteria. By combining various search terms related to ML, safety analysis, and V\&V identified during the project definition phase of the workshop series (cf. a) in Fig~\ref{fig:processOverview}), we identified 14 papers representing a diversity of authors, publishers, and publications venues, i.e., adhering to recommendations for a feasible start set\cite{wohlin_guidelines_2014}. Still, the composition of the start set is a major threat to the validity of any snowballing literature review. %Consequently, we used experiences collected during the project definition phase of the workshop series to seed the literature review. 
Table~\ref{tab:snowball} shows the papers in the start set.

Originating in the 14 papers in the start set, we iteratively conducted backward and forward snowballing. Backward snowballing means scanning the reference lists for additional papers to include. Forward snowballing from a paper involves adding related papers that cite the given paper. We refer to one combined effort of backward and forward snowballing as an \textit{iteration}. In each iteration, two researchers collected candidates for inclusion and two other researchers validated the selection using the inclusion criteria. Despite our efforts to carefully process iterations, there is always a risk that relevant publications could not be identified by following references from our start set due to citation patterns in the body of scientific literature, e.g., research cliques.

\subsubsection{Data extraction and analysis}
When the snowballing was completed, two authors extracted publication metadata according to a pre-defined extraction form, e.g., publication venue and application domain. Second, the same two authors conducted an assessment of rigor and relevance as recommended by Ivarsson and Gorschek\cite{ivarsson_method_2011}. Third, they addressed RQ1 using thematic analysis\cite{cruzes_recommended_2011}, i.e., summarizing, integrating, combining, and comparing findings of primary studies to identify patterns. %Consequently, we extracted patterns regarding the state-of-the-art of V\&V for ML-based systems, including both challenges and solutions proposals. 

Our initial plan was to classify challenges and solution proposals in previous work using classification schemes developed by Amodei \textit{et al.} \textbf{[P2]} and Varshney\cite{varshney_engineering_2016}, respectively. However, neither of the two proposed categorization schemes were successful in spanning the content of the selected papers. To better characterize the selected body of research, we inductively created new classification schemes for challenges and solution proposals according to a grounded theory approach. Table~\ref{tab:cat} defines the final categories used in our study, seven challenge categories and five solution proposal categories.

\begin{table*}[]
\centering
\caption{Definition of categories of challenges and solution proposals for V\&V of ML-based systems.}
\label{tab:cat}
\begin{tabular}{|l|p{12cm}|}
\hline
\textbf{Challenge Categories}         & \textbf{Definitions}                                                                                 \\ \hline
State-space explosion                 & Challenges related to the very large size of the input space.                                        \\ \hline
Robustness                            & Issues related to operation in the presence of invalid inputs or stressful environmental conditions. \\ \hline
Systems engineering                   & Challenges related to integration or co-engineering of ML-based and conventional components.         \\ \hline
Transparency                          & Challenges originating in the black-box nature of the ML system.                                     \\ \hline
Requirements specification            & Problems related to specifying expectations on the learning behavior.                                \\ \hline
Test specification                    & Issues related to designing test cases for ML-based systems, e.g., non-deterministic output.         \\ \hline
Adversarial attacks                   & Threats related to antagonistic attacks on ML-based systems, e.g., adversarial examples.             \\ \hline
                                      &                                                                                                      \\ \hline
\textbf{Solution Proposal Categories} & \textbf{Definitions}                                                                                 \\ \hline
Formal methods                        & Approaches to mathematically prove that some specification holds.                                    \\ \hline
Control theory                        & Verification of learning behavior based on automatic control and self-adaptive systems.              \\ \hline
Probabilistic methods                 & Statistical approaches such as uncertainty calculation, Bayesian analysis, and confidence intervals. \\ \hline
Test case design                      & Approaches to create effective test cases, e.g., using genetic algorithms or procedural generation.  \\ \hline
Process guidelines                    & Guidelines supporting work processes, e.g., covering training data collection or testing strategies. \\ \hline
\end{tabular}
\end{table*}

\subsection{The questionnaire-based survey}
To validate the findings from the snowballed literature (cf. C. in Figure~\ref{fig:evidence}), we designed a web-based questionnaire to survey practitioners in safety-critical domains. Furthermore, reaching out to additional practitioners beyond the SMILE project enables us to collect more insights into challenges related to ML-based systems in additional safety-critical contexts (cf. D. in Fig.~\ref{fig:evidence}). Moreover, we used the survey to let the practitioners rate the importance of the challenges reported in the academic literature, as well as the perceived feasibility of the published solutions proposals.

We designed the survey instrument using Google Forms, structured as 10 questions organized into two sections. The first section consisted of seven closed-end questions related to demographics of the respondents and their organizations and three Likert items concerning high-level statements on V\&V of ML-based systems. The second section consisted of three questions: 1) rating the importance of the challenge categories, 2) rating how promising the solution proposal categories are, and 3) an open-end free-text answer requesting a comment on our main findings and possibly adding missing aspects. 

We opted for an inclusive approach and used convenience sampling to collect responses\cite{rea_designing_2014}, i.e., a non-probabilistic sampling method. The target population was software and systems engineering practitioners working in safety-critical contexts, including both engineering and managerial roles, e.g., test managers, developers, architects, safety engineers, and product managers. The main recruitment strategy was to invite the extended SMILE network (cf. workshops \#5 and \#6 in Fig.~\ref{fig:processOverview}) and to advertise the survey invitation in LinkedIn groups related to development of safety-critical systems. %We posted the invitation in 20 LinkedIn groups, e.g., ``ISO 26262 For Engineers \& Managers'', ``Safety Critical Professionals'', and ``Automotive Testing''. 
We collected answers in 2017, from July 1 to August 31. 

As a first step of the response analysis, we performed a content sanity check to identify invalid answers, e.g., nonsense or careless responses. Subsequently, we collected summary statistics of the responses and visualized it with bar charts to get a quick overview of the data. We calculated Spearman rank correlation ($\rho$) between all ordinal scale responses, interpreting correlations as weak, moderate, and strong for $\rho>0.3$, $\rho>0.5$, and $\rho>0.7$, respectively. Finally, the two open-ended questions were coded, summarized, and validated by four of the co-authors.

\section{Results and Discussion} \label{sec:res}
This section is organized according to the evidence perspective provided in Fig.~\ref{fig:evidence}: A. Related work, B. Workshop findings, C. Snowballed literature, and D) Survey responses. As reported in Section~\ref{sec:method}, A. and B. focus on industrial relevance, whereas C. and D. aim at academic rigor.

\subsection{Related work}
The related work section (cf. A. in Fig.~\ref{fig:evidence}) presents an overview of literature that was identified during the SMILE project. Fourteen of the papers were selected early to seed the (independent) snowballing literature review described in Section~\ref{sec:slr}. In this section, we first describe the start set \textbf{[P1]}-\textbf{[P14]}, and then papers that were subsequently identified by SMILE members or the anonymous reviewers of the manuscript -- but not through the snowballing process (as these are reported separately in Section~\ref{sec:snowball}).

\subsubsection{The snowballing start set} \label{sec:seed}
The following 14 papers were selected as the snowballing start set, representing a diverse set of authors, publication venues, and publication years. We briefly describe them below, and motivate their inclusion in the start set.
\begin{itemize}
\item[\textbf{[P1]}] Clark \textit{et al.} reported from a US Air Force research project on challenges in V\&V of autonomous systems. This work is highly related to the SMILE project.
\item[\textbf{[P2]}] Amodei \textit{et al.} listed five challenges to artificial intelligence safety according to Google Brain: 1) avoiding negative side effects, 2) avoiding reward hacking, 3) scalable oversight, 4) safe exploration, and 5) robustness to distributional shift.
\item[\textbf{[P3]}] Brat and Jonsson discussed challenges in V\&V of autonomous systems engineered for space exploration. Included to cover the space domain.
\item[\textbf{[P4]}] Broggi \textit{et al.} presented extensive testing of the BRAiVE autonomous vehicle prototype by driving from Italy to China. Included as it is different, i.e., reporting experiences from a practical trip.
\item[\textbf{[P5]}] Taylor \textit{et al.} sampled research in progress (in 2003) on V\&V of neural networks, aimed at NASA applications. Included to snowball research conducted in the beginning of the millennium.
\item[\textbf{[P6]}] Taylor \textit{et al.} with the Machine Intelligence Research Institute surveyed design principles that could ensure that systems behave in line with the interests of their operators -- which they refer to as ``AI alignment''. Included to bring in a more philosophical perspective on safety.
\item[\textbf{[P7]}] Carvalho \textit{et al.} presented a decade of research on control design methods for systematic handling of uncertain forecasts for autonomous vehicles. Included to cover robotics.
\item[\textbf{[P8]}] Ramos \textit{et al.} proposed a DNN-based obstacle detection framework, providing sensor fusion for detection of small road hazards. Included as the work closely resembles the use case discussed at the workshops (see Section~\ref{sec:ws}).
\item[\textbf{[P9]}] Alexander \textit{et al.} suggested ``situation coverage methods'' for autonomous robots to support testing of all environmental circumstances. Included to cover coverage.
\item[\textbf{[P10]}] Zou \textit{et al.} discussed safety assessments of probabilistic airborne collision avoidance systems and proposes a genetic algorithm to search for undesired situations. Included to cover probabilistic approaches.
\item[\textbf{[P11]}] Zou \textit{et al.} presented a safety validation approach for avoidance systems in unmanned aerial vehicles, using evolutionary search to guide simulations to potential conflict situations in large state spaces. Although the authors overlap, included to snowball research on simulation.
\item[\textbf{[P12]}] Arnold and Alexander proposed using procedural content generation to create challenging environmental situations when testing autonomous robot control algorithms in simulations. Included to cover synthetic test data.
\item[\textbf{[P13]}] Sivaraman and Trivedi compared three active learning approaches for on-road vehicle detection. Included to add a semi-supervised ML approach.
\item[\textbf{[P14]}] Mozaffari \textit{et al.} developed a robust safety-oriented autonomous cruise controller based on the model predictive control technique. Included to identify approaches based on control theory. 
\end{itemize}

In the start set, we consider \textbf{[P1]} to be the research endeavor closest to our current study. While we target the automotive domain rather than aerospace, both studies address highly similar research objectives -- and also the method used to explore the topic is close to our approach. \textbf{[P1]} describes a year-long study aimed at: 1) understanding the unique challenges to the certification of safety-critical autonomous systems and 2) identifying the V\&V approaches needed to overcome them. To accomplish this, the US Air Force organized three workshops with representatives from industry, academia, and governmental agencies, respectively.
\textbf{[P1]} concludes that that there are four enduring problems that must be addressed:
\begin{itemize}
\item State-Space Explosion -- In an autonomous system, the decision space is non-deterministic and the system might be continuously learning. Thus, over time, there may be several output signals for each input signal. This in turn makes it inherently challenging to exhaustively search, examine, and test the entire decision space.
\item Unpredictable Environments -- Conventional systems have limited ability to adapt to unanticipated events, but an autonomous systems should respond to situations that were not programmed at design time. However, there is a trade-off between performance and correct behavior, which exacerbates the state-space explosion problem.  
\item Emergent Behavior -- Non-deterministic and adaptive systems may induce behavior that result in unintended consequences. Challenges comprise how to understand all intended and unintended behavior and how to design experiments and test vectors that are applicable to adaptive decision making in an unpredictable environment. 
\item Human-Machine Communication -- Hand-off, communication, and cooperation between the operator and the autonomous system play an important role to create mutual trust between the human and the system. It is not known how to address these issues when the behavior is not known at design time. 
\end{itemize}

With these enduring challenges in mind, \textbf{[P1]} calls for research to pursue five goals in future technology development. First, approaches to \textit{cumulatively build safety evidence} through the phases of Research \& Development (R\&D), Test \& Evaluation (T\&E), and Operational Tests. The US Air Force calls for effective methods to reuse safety evidence throughout the entire product development lifecycle. Second, \textbf{[P1]} argues that \textit{formal methods}, embedded during R\&D, could provide safety assurance. This approach could reduce the need for T\&E and operational tests. Third, novel techniques to \textit{specify requirements based on formalism, mathematics, and rigorous natural language} could bring clarity and allow automatic test case generation and automated traceability to low-level designs. Fourth, \textit{run-time decision assurance} may allow restraining the behavior of the system, thus shifting focus from off-line verification to instead performing on-line testing at run-time. Fifth, \textbf{[P1]} calls for research on \textit{compositional case generation}, i.e., better approaches to combine different pieces of evidence into one compelling safety case.

\subsubsection{Non-snowballed related work}
This subsection reports the related work that stirred up the most interesting discussions in the SMILE project. In contrast to the snowballing literature review, we do not provide steps to replicate the identification of the following papers.

Knauss \textit{et al.} conducted an exploratory interview study to elicit challenges when engineering autonomous cars\cite{knauss_software-related_2017}. Based on interviews and focus groups with 26 domain experts in five countries, the authors report in particular challenges in testing automated vehicles. Major challenges are related to: 1) virtual testing and simulation, 2) safety, reliability, and quality, 3) sensors and their models 4) complexity of, and amount of, test cases, and 5) hand-off between driver and vehicle.

Spanfelner \textit{et al.} conducted research on safety and autonomy in the ISO~26262 context\cite{spanfelner_challenges_2012}. Their conclusion is that driver assistance systems need models to be able to interpret the surrounding environment, i.e., to enable vehicular perception. Since models, by definition, are simplifications of the real world, they will be subject to functional insufficiencies. By accepting that such insufficiencies may fail to reach the functional safety goals, it is possible to design additional measures that in turn can meet the safety goals.

Heckemann \textit{et al.} identified two primary challenges in developing autonomous vehicles adhering to ISO 26262\cite{heckemann_safe_2011}. First, the driver is today considered to be part of the safety concept, but future vehicles will make driving maneuvers without interventions by a human driver. Second, the system complexity of modern vehicle systems is continuously growing as new functionality is added. This obstructs safety assessment, as increased complexity makes it harder to verify freedom of faults.

Varshney discussed concepts related to engineering safety for ML systems from the perspective of minimizing risk and epistemic uncertainty\cite{varshney_engineering_2016}, i.e., uncertainty due to gaps in knowledge as opposed to intrinsic variability in the products. 
More specifically, he analyzed how four general strategies for promoting safety\cite{moller_principles_2008} apply to systems with ML components. First, \textit{inherently safe design} means excluding a potential hazard from the system instead of controlling it. A prerequisite for assuring such a design is to improve the interpretability of the typically opaque ML models. Second, \textit{safety reserves} means the factor of safety, e.g., the ratio of absolute structural capacity to actual applied load in structural engineering. In ML, interpretations include a focus on a the maximum error of classifiers instead of the average error, or training models to be robust to adversarial examples. Third, \textit{safe fail} implies that a system remains safe even when it fails in its intended operation, traditionally by relying on constructs such as electrical fuses and safety valves. In ML, a concept of run-time monitoring must be accomplished, e.g., by continuously monitoring how certain a DNN model is performing in its classification task. Fourth, \textit{procedural safeguards} covers any safety measures that are not designed into the system, e.g., mandatory safety audits, training of personnel, and user manuals describing how to define the training set. 

Seshia \textit{et al.} identified five major challenges to achieve formally-verified AI-based systems \cite{seshia_towards_2016}. First, a methodology to provide a model of the environment even in the presence of uncertainty. Second, a precise mathematical formulation of what the system is supposed to do, i.e., a formal specification. Third, the need to come up with new techniques to formally model the different components that will use machine learning. Fourth, systematically generating training and testing data for ML-based components. Finally, developing computationally scalable engines that are able to verify quantitatively the requirements of a system.

One approach to tackle the opaqueness of DNNs is to use visualization. Bojarski \textit{et al.}\cite{bojarski_visualbackprop:_2016} developed a tool for visualizing the parts of an image that are used for decision making in vehicular perception. Their tool demonstrated an end-to-end driving application where the input is images and the output is the steering angle. Mhamdi \textit{et al.} also studied the black box aspects of neural networks, and show that the robustness of a complete DNN can be assessed by an analysis focused on individual neurons as units of failure\cite{mhamdi_robustness_2017} -- a much more reasonable approach given the state-space explosion.

In a paper on ensemble learning, Varshney \textit{et al.} describes a reject option for classifiers\cite{varshney_practical_2013}. Such a classifier could, instead of presenting a highly uncertain classification, request that a human operator must intervene. A common assumption is that the classifier is the least confident in the vicinity of the decision boundary, i.e., that there is an inverse relationship between distance and confidence. While this might be true in some parts of the feature space, it is not a reliable measure in parts that contain too few training examples. For a reject option to provide a ``safe fail'' strategy, it must trigger both 1) near the decision boundary in parts of the feature space with many training examples, and 2) in any decision represented by too few training examples.

Heckemann \textit{et al.} proposed using the concept of \textit{adaptive safety cage architectures} to support future autonomy in the automotive domain\cite{heckemann_safe_2011}, i.e., an independent safety mechanism that continuously monitors sensor input. The authors separated two areas of operation: a valid area (that is considered safe) and an invalid area that can lead to hazardous situations. If the function is about to enter the invalid area, the safety cage will invoke an appropriate safe action, such as a minimum risk emergency stopping maneuver or a graceful degradation. Heckemann \textit{et al.} argued that a safety cage can be used in an ASIL decomposition by acting as a functionally redundant system to the actual control system. The highly complex control function could then be developed according to the quality management standard, whereas the comparably simple safety cage could adhere to a higher ASIL level.

Adler \textit{et al.} presented a similar run-time monitoring mechanism for detecting malfunctions, referred to as a \textit{safety supervisor}\cite{adler_safety_2016}. Their safety supervisor is part of an overall safety approach for autonomous vehicles, consisting of a structured four-step method to identify the most critical combinations of behaviors and situations. Once the critical combinations have been specified, the authors propose implementing tailored safety supervisors to safeguard against related malfunctions.

Finally, a technical report prepared by Bhattacharyya \textit{et al.} for the NASA Langley Research Center discussed certification considerations of adaptive systems in the aerospace domain\cite{bhattacharyya_certification_2015}. The report separates adaptive control algorithms and Artificial Intelligence (AI) algorithms, and the latter is closely related to our study since it covers machine learning and ANN.  Their certification challenges for adaptive systems are organized in four categories:

\begin{itemize}
\item Comprehensive requirements -- Specifying a set of requirements that completely describe the behavior, as mandated by current safety standards, is presented as the most difficult challenge to tackle.
\item Verifiable requirements -- Specifying pass criteria for test cases at design-time might be hard. Also, current aerospace V\&V relies heavily on coverage testing of source code in imperative languages, but how to interpret that for AI algorithms is unclear.
\item Documented design -- Certification requires detailed documentation, but components realizing adaptive algorithms were rarely developed with this in mind. Especially AI algorithms are often distributively developed by open source communities, which makes it hard to reverse engineer documentation and traceability.
\item Transparent design -- Regulators expect a transparent design and a conventional implementation to be presented for evaluation. Increasing system complexity by introducing novel adaptive algorithms challenges comprehensibility and trust. On top of that, adaptive systems are often non-deterministic, which makes it harder to demonstrate absence of unintended functionality.
\end{itemize}

\subsection{The Workshop Series} \label{sec:ws}
During the six workshops with industry partners (cf. \#1-\#6 in Fig.~\ref{fig:processOverview}), we discussed key questions that must be explored to enable engineering of safety-critical automotive systems with DNNs. %We elicited the most pressing challenges, and discussed various solution proposals in an open manner. 
Three sub-areas emerged during the workshops: 1) robustness, 2) interplay between DNN components and conventional software, and 3) V\&V of DNN components.

\subsubsection{Robustness of DNN Components}
The concept of robustness permeated most discussions during the workshops. While robustness is technically well-defined, in the workshops it often remained a rather elusive quality attribute -- typically translated to ``something you can trust''. 

To bring the workshop participants to the same page, we found it useful to base the discussions on a simple ML case: a confusion matrix for a one-class classifier for camera-based animal detection. For each input image, the result of the classifier is limited to one of the four options: 1) an animal is present and correctly classified (true positive), 2) no animal is present and the classifier does not signal animal detection (true negative), 3) the classifier reports animal presence, but there is none (false positive), and 4) an animal is present, but the classifier misses it (false negative).

For the classifier to be considered robust, the participants stressed the importance of not generating false positives and false negatives despite occasional low quality input or changes in the environmental conditions, e.g., dusk, rain, or sun glare. A robust ML system should neither miss present animals, risking collisions, nor suggest emergency braking that risk rear-end collisions. As the importance of robustness in the example is obvious, we see a need for future research both on how to specify and verify acceptable levels of ML robustness.

During the workshops, we also discussed more technical aspects engineering robust DNN components. First, our industry practitioners brought up the issue of DNN architectures to be problem-specific. While there are some approaches to automatically generating neural network architectures\cite{angeline_evolutionary_1994,yao_new_1997}, typically designing the DNN architecture is an \textit{ad hoc} process of trial and error. Often a well known architecture is used as a baseline and then it is tuned to fit the problem at hand. Our workshops recognized the challenge of engineering robust DNN-based systems, in part due to their highly problem-specific architectures. 

Second, once the DNN architecture is set, training commences to assign weights to the trainable parameters of the network. The selection of training data must be representative for the task, in our discussions animal detection, and for the environment that the system will operate in. The workshops agreed that robustness of DNN components can never be achieved without careful selection of training data. Not only must the amount and quality of sensors (in our case cameras) acquiring the different stimuli for the training data be sufficient, also other factors such as positioning, orientation, aperture, and even geographical location like city and country must match the animal detection example. At the workshops, we emphasized the issue of camera positions as both car and truck manufacturers were part of SMILE -- to what extent can training data from a car's perspective be reused for a truck? Or should a truck rather benefit from its size and collect dedicated training data from its elevated camera position?  %Thus, even though initial data sets are available, they represent only a small part of the scenarios that may occur on a regular road.

Third, also related to training data, the workshops discussed working with synthetic data. While such data always can be used to complement training data, there are several open questions on how to best come up with the best mix during the training stage. As reported in Section~\ref{sec:dnn}, GANs\cite{goodfellow_nips_2016,radford_unsupervised_2015} could be a good tool for synthesizing data. Sixt~\textit{et al.}\cite{sixt_rendergan:_2016} proposed a framework called RenderGAN that could generate large amounts of realistic labeled data for training. In transfer learning, training efficiency improves by combining data from different data sets \cite{glorot_domain_2011,oquab_learning_2014}. One possible approach could be to first train the DNN component using synthetic data from, e.g., simulators like TORCS\footnote{http://torcs.sourceforge.net}, then data from some publicly available database could be used to continue the training, e.g., the KITTI data\footnote{http://www.cvlibs.net/datasets/kitti/} or CityScape\footnote{https://www.cityscapes-dataset.com/}, and finally, data from the geographical region where the vehicle should operate could be added. For any attempts at transfer learning, the workshops identified the need to measure to what extent training data matches the planned operational environment.

\subsubsection{Complementing DNNs with Conventional Components}
During the workshops, we repeatedly reminded the participants to consider DNNs from a systems perspective. DNN components will always be part of an automotive system consisting of also conventional hardware and software components.

Several researchers claim that that DNN components is a prerequisite for autonomous driving\cite{falcini_deep_2017,agrawal_accelerator_2017,sallab_end--end_2016}. However, how to integrate such components in a system is an open question. Safety is a systems issue, rather than a component specific issue. All hazards introduced by both DNNs and conventional software must be analyzed within the context of systems engineering principles. On the other hand, the hazards can also be addressed on a system level.

One approach to achieve DNN safety is to introduce complementary components, i.e., when a DNN model fails to generalize, a conventional software or hardware component might step in to maintain safe operation. During the workshops, particular attention was given to introducing a \textit{safety cage concept}. Our discussions orbited a solution in which the DNN component was encapsulated by a supervisor, or a safety cage, that continuously monitors the input to the DNN component. The envisioned safety cage should perform novelty detection~\cite{pimentel_review_2014} and alert when input does not belong within the training region of the DNN component, i.e., if the risk of failed generalization was too high, the safety cage should re-direct the execution to a \textit{safe-track}. The safe-track should then operate  without any ML components involved, enabling traditional approaches to safety-critical software engineering. %Such a concept has previously been applied within other domains\cite{englund_som-based_2007,englund_combining_2007}, and the concept was also described by Varshney\cite{varshney_engineering_2016}.  

The concept of an ML safety cage is in line with Varshney's discussions of ``safe fail''\cite{varshney_engineering_2016}. Different options to implement an ML safety cage include adaptations of fail-silent systems\cite{brasileiro_implementing_1996}, plausibility checks \cite{korte_design_2012}, and arbitration. However, Adler~\textit{et al.}\cite{adler_safety_2016} indicated that the \emph{no free lunch} theorem might apply for safety cages, by stating that if tailored safety safety cages are to be developed to safeguard against domain-specific malfunctions, thus, different safety cages may be required for different systems. 

Introducing redundancy in the ML system is an approach related to the safe track. One method is to use ensemble methods in computer vision applications\cite{maji_ensemble_2016}, i.e., employing multiple learning algorithms to improve predictive performance. Redundancy can also be introduced in an ML-based system using hardware component, e.g., using an array of sensors of the same, or different, kind. Increasing the amount of input data should increase the probability of finding patterns closer to the training data set. Combining data from various input sources, referred to as sensor fusion, also helps overcoming the potential deficiencies of individual sensors.

\subsubsection{V\&V Approaches for Systems with DNN Components}
Developing approaches to engineer robust systems with DNN components is not enough, the automotive industry must also develop novel approaches to V\&V. 
V\&V is a cornerstone in safety certification, but it still remains unclear how to develop a safety case around applications with DNNs.%This subsection summarizes related industry perspectives captured during the workshops. 

As pointed out in previous work, the current ISO~26262 standard is not applicable when developing autonomous systems that rely on DNNs\cite{heckemann_safe_2011}. Our workshops corroborate this view, by identifying several open questions that need to be better understood:
\begin{itemize}
\item How is a DNN component classified in ISO~26262? Should it be regarded as an individual software unit or a component?
\item From a safety perspective, is it possible to treat DNN misclassifications as ``hardware failures''? If yes, are the hardware failure target values defined in ISO~26262 applicable?
\item ISO~26262 mandates complete test coverage of the software, but what does this imply for a DNN? What is sufficient coverage for a DNN?
\item What metrics should be used to specify the DNN accuracy? Should quality targets using such metrics be used in the DNN requirements specifications, and subsequently as targets for verification activities?
\end{itemize}

Apart from the open questions, our workshop participants identified several aspects that would support V\&V. First, as requirements engineering is fundamental to high-quality V\&V\cite{bjarnason_challenges_2014}, some workshop participants requested a formal, or semi-formal, notation for requirements related to functional safety in the DNN context. Defining low-level requirements that would be verifiable appears to be one of the greatest challenges in this area. Second, there is a need for a tool-chain and framework tailored to lifecycle management of systems with DNN components -- current solutions tailored for human-readable source code are not feasible and must be complemented with too many immature internal tools. Third, methods for test case generation for DNN will be critical, as manual creation of test data does not scale.

Finally, a major theme during the workshops was how to best use simulation as a means to support V\&V. We believe that the future will require massive use of simulation to ensure safe DNN components. Consequently, there is a need to develop simulation strategies to cover both normal circumstances as well as rare, but dangerous, traffic situations. Furthermore, simulation might also be used to assess the sensitivity to adversarial examples. 

\subsection{The systematic snowballing} \label{sec:snowball}
Table~\ref{tab:snowball} shows the results from the five iterations of the snowballing. In total, the snowballing procedure identified 64 papers including the start set. We notice two publication peaks: 29 papers were published between 2002-2007 and 25 papers were published between 2013-2016. The former set of papers were dominated by research on using neural networks for adaptive flight controllers, whereas the latter set predominantly addresses the automotive domain. This finding suggests that organizations currently developing ML-based systems for self-driving cars could learn from similar endeavors in the aerospace domain roughly a decade ago -- while DNN was not available then, several aspects of V\&V enforced by aerospace safety standards are similar to ISO~26262. Note, however, that 19 of the papers do not target any specific domain, but rather discusses ML-based systems in general.

Table~\ref{tab:chalsol} shows the distribution of challenge and solution proposal categories identified in the papers; `\#' indicates the number of unique challenges or solution proposals matching a specific category. As each paper can report more than one challenge or solution proposal, and the same challenge or solution proposal can occur in more than one paper, the number of Paper IDs in the third column does not necessarily match the `\#'. The challenges most frequently mentioned in the papers relate to state-space explosion and robustness, whereas the most commonly proposed solutions constitute approaches that belong to formal methods, control theory, or probabilistic methods. 

\begin{table*}[]
\centering
\tcap{Distribution of challenge and solution proposal categories.}
\label{tab:chalsol}
\begin{tabular}{lll}
\hline
\multicolumn{1}{|l|}{\textbf{Challenge category}}                  & \multicolumn{1}{l|}{\textbf{\#}} & \multicolumn{1}{l|}{\textbf{Paper IDs}}                                                                                                                                    \\ \hline
\multicolumn{1}{|l|}{State-space explosion}      & \multicolumn{1}{l|}{6}  & \multicolumn{1}{l|}{{[}P3{]}, {[}P15{]}, {[}P16{]}, {[}P47{]}}                                                                                                    \\ \hline
\multicolumn{1}{|l|}{Robustness}                 & \multicolumn{1}{l|}{4}  & \multicolumn{1}{l|}{{[}P1{]}, {[}P2{]}, {[}P15{]}, {[}P55{]}}                                                                                                     \\ \hline
\multicolumn{1}{|l|}{Systems engineering}        & \multicolumn{1}{l|}{2}  & \multicolumn{1}{l|}{{[}P1{]}, {[}P55{]}}                                                                                                                          \\ \hline
\multicolumn{1}{|l|}{Transparency}               & \multicolumn{1}{l|}{2}  & \multicolumn{1}{l|}{{[}P1{]}, P55{]}}                                                                                                                             \\ \hline
\multicolumn{1}{|l|}{Requirements specification} & \multicolumn{1}{l|}{3}  & \multicolumn{1}{l|}{{[}P15{]}, {[}P55{]}}                                                                                                                         \\ \hline
\multicolumn{1}{|l|}{Test specification}         & \multicolumn{1}{l|}{3}  & \multicolumn{1}{l|}{{[}P16{]}, {[}P46{]}, {[}P55{]}}                                                                                                              \\ \hline
\multicolumn{1}{|l|}{Adversarial attacks}        & \multicolumn{1}{l|}{1}  & \multicolumn{1}{l|}{{[}P15{]}}                                                                                                                                    \\ \hline
                                                 &                         &                                                                                                                                                                   \\ \hline
\multicolumn{1}{|l|}{\textbf{Solution proposal category}}          & \multicolumn{1}{l|}{\textbf{\#}} & \multicolumn{1}{l|}{\textbf{Paper IDs}}                                                                                                                                    \\ \hline
\multicolumn{1}{|l|}{Formal methods}             & \multicolumn{1}{l|}{8}  & \multicolumn{1}{l|}{\begin{tabular}[c]{@{}l@{}}{[}P3{]}, {[}P26{]}, {[}P42{]}, {[}P28{]}, \\ {[}P37{]}, P{[}40{]}, {[}P44{]}, {[}P53{]}\end{tabular}}             \\ \hline
\multicolumn{1}{|l|}{Control theory}             & \multicolumn{1}{l|}{7}  & \multicolumn{1}{l|}{\begin{tabular}[c]{@{}l@{}}{[}P7{]}, {[}P20{]}, {[}P25{]}, {[}P64{]}, \\ {[}P36{]}, {[}P47{]}, {[}P57{]}, {[}P60{]}\end{tabular}}             \\ \hline
\multicolumn{1}{|l|}{Probabilistic methods}      & \multicolumn{1}{l|}{7}  & \multicolumn{1}{l|}{\begin{tabular}[c]{@{}l@{}}{[}P18{]}, {[}P30{]}, {[}P31{]}, {[}P32{]}, {[}P33{]}, \\ {[}P35{]}, {[}P50{]}, {[}P52{]}, {[}P54{]}\end{tabular}} \\ \hline
\multicolumn{1}{|l|}{Test case design}           & \multicolumn{1}{l|}{5}  & \multicolumn{1}{l|}{{[}P9{]}, {[}P10{]}, {[}P12{]}, {[}P17{]}, {[}P21{]}}                                                                                         \\ \hline
\multicolumn{1}{|l|}{Process guidelines}         & \multicolumn{1}{l|}{4}  & \multicolumn{1}{l|}{{[}P23{]}, {[}P51{]}, {[}P56{]}, {[}P59{]}}                                                                                                   \\ \hline
\end{tabular}
\end{table*}

Regarding the publication years, we notice that the discussion on state-space explosion primarily has been active in recent years, possibly explained by the increasing application of DNNs. Looking at solution proposals, we see that probabilistic methods was particularly popular during the first publication peak, and that research specifically addressing test case design for ML-based systems has appeared first after 2012.  

Fig.~\ref{fig:mapping} shows a mapping between solution proposals categories and challenge categories. Some of the papers propose a solution to address challenges belonging to a specific category. For each such instance, we connect solution proposals (to the left) and challenges (to the right), i.e., the width of the connection illustrates the number of instances. Note that we did put the solution proposal in \textbf{[P4]} (deployment in real operational setting) in its own `Other' category. None of the proposed solutions address challenges related to the categories ``Requirements specification'' or ``Systems engineering'', indicating a research gap. Furthermore, ``Transparency'' is the challenge category that has been addressed the most in the papers, followed by ``State-space explosion''.

\begin{Figure}
\centering
\includegraphics[width=0.99\textwidth]{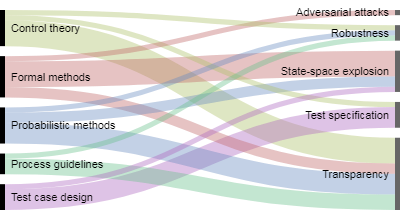}
\fcaption{Mapping between categories of solution proposals (to the left) and challenges (to the right).}
\label{fig:mapping}
\end{Figure}

Two books summarize most findings from the aerospace domain identified through our systematic snowballing. Taylor edited a book in 2006 that collected experiences for V\&V of ANN technology\cite{taylor_methods_2006} in a project sponsored by the NASA Goddard Space Flight Center. Taylor concluded that the V\&V techniques available at the time must evolve to tackle ANNs. Taylor's book reports five areas that need to be augmented to allow V\&V of ANN-based systems:\footnote{The best practices were also later distilled into a guidance document intended for practitioners\cite{pullum_guidance_2007}}
\begin{itemize}
\item \textit{Configuration management} must track all additional design elements, e.g., the training data, the network architecture, and the learning algorithms. Any V\&V activity must carefully specify the configuration under test. 
\item \textit{Requirements} need to specify novel adaptive behavior, including control requirements (how to acquire and act on knowledge) and knowledge requirements (what knowledge should be acquired).
\item \textit{Design specifications} must capture design choices related to novel design elements such as training data, network architecture, and activation functions. V\&V of the ANN design should ensure that the choices are appropriate.
\item \textit{Development lifecycles} for ANNs are highly iterative and last until some quantitative goal has been reached. Traditional waterfall software development is not feasible, and V\&V must be an integral part rather than an add-on.
\item \textit{Testing} needs to evolve to address novel requirements. Structure testing should determine whether the network architecture is better at learning according to the control requirements than alternative architectures. Knowledge testing should verify that the ANN has learned what was specified in the knowledge requirements.
\end{itemize}

The second book that has collected experiences on V\&V of (mostly aerospace) ANNs, also funded by NASA, was edited by Schumann and Liu and published in 2010\cite{schumann_application_2010}. While the book primarily surveys the use of ANNs in high-assurance systems, parts of the discussion is focused on V\&V -- and the overall conclusion that V\&V must evolve to handle ANNs is corroborated. In contrast to the organization we report in Table~\ref{tab:chalsol}, the book suggests grouping solution proposals into approaches that: 1) separate ANN algorithms from conventional source code, 2) analyze the network architecture, 3) consider ANNs as function approximators, 4) tackle the opaqueness of ANNs, 5) assess the characteristics of the learning algorithm, 6) analyze the selection and quality of training data, and 7) provides means for online monitoring of ANNs. We believe that our organization is largely orthogonal to the list above, thus both could be used in a complementary fashion.

\subsection{The survey}
This section organizes the findings from the survey into closed questions, correlation analysis, and open questions, respectively.

\subsubsection{Closed questions}
Forty-nine practitioners answered our survey, most of them primarily working in Europe (38 out of 49, 77.6\%). Twenty respondents (40.8\%) work primarily in the automotive domain, followed by 14 in aerospace (28.6\%). Other represented domains include process industry (5 respondents), railway (5 respondents), and government/military (3 respondents). The respondents represent a variety of roles, from system architects (17 out of 49, 34.7\%) to product developers (10 out of 49, 20.4\%), and managerial roles (7 out of 49, 14.3\%). Most respondents primarily work in Europe (38 out of 49, 77.6\%) or North America (7 out of 49, 14.3\%).%Further details about the background of the respondents, and all other questions, are presented in Appendix (TODO).

Most respondents have some proficiency in ML. Twenty-five respondents (51.0\%) report having fundamental awareness of ML concepts and practical ML concerns. Sixteen respondents (32.7\%) have higher proficiency, i.e., can implement ML solutions independently or with guidance -- but no respondents consider themselves ML experts. On the other side of the spectrum, eight respondents report possessing no ML knowledge.

We used three Likert items to assess the respondents' general thoughts about ML and functional safety, reported as a)-c) in Table~\ref{tab:survey}. Most respondents agree (or strongly agree) that applying ML in safety-critical applications will be important in their organizations in the future (29 out of 49, 59.2\%), whereas eight (16.3\%) disagree. At the same time, 29 out of 49 (59.2\%) of the respondents report that V\&V of ML-based features is considered particularly difficult by their organizations -- 20 respondents even strongly agrees with the statement. It is clear to our respondents that more attention is needed regarding V\&V of ML-based systems, as only 10 out of 49 (20.4\%) believe that their organizations are well-prepared for the emerging paradigm.

\begin{table*}
\centering
\tcap{Answers to the closed questions of the survey. a)-c) show three Likert items, ranging from strongly disagree (1) to strongly agree (5). d)- o) reports on importance/promisingness using the following ordinal scale: not at all, slightly, somewhat, moderately, and extremely. The `Missing`'' column includes both ``I don't know'' answers and missing answers.}
\includegraphics[width=0.85\textwidth]{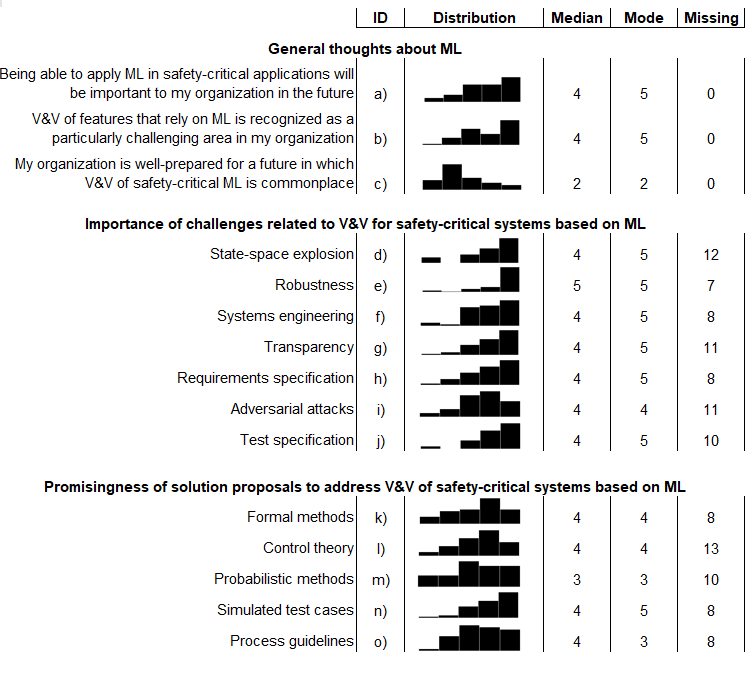}
\label{tab:survey}
\end{table*}

Robustness (cf. e) in Table~\ref{tab:survey}) stands out as the particularly important challenge, reported as ``extremely important'' by 29 out of 49 (59.2\%). However, all challenges covered in the questionnaire were considered important by the respondents. The only challenge that appears less urgent to the respondents is adversarial attacks, but the difference is minor.

The respondents consider simulated test cases as the most promising solution proposal to tackle challenges in V\&V of ML-based systems, reported as extremely promising by 18 out of 49 respondents (36.7\%) and moderately promising by 12 respondents (24.5\%). Probabilistic methods is the least promising solution proposal according to the respondents, followed by process guidelines.

\subsubsection{Correlation analysis}

We identified some noteworthy correlations in the responses. The respondents' ML proficiency (Q4) is moderately correlated ($\rho=0.53$) with the perception of ML importance (Q5) -- an expected finding as respondents with a personal investment are likely to be biased. More interestingly, we found that ML proficiency was also moderately correlated to two of the seven challenge categories: transparency ($\rho=0.61$) and state-space explosion ($\rho=0.54$). This suggests that these two challenges are particularly difficult to comprehend for non-experts. Perceiving the organization as well-prepared for introducing ML-based solutions (Q4) is moderately correlated ($\rho=0.57$) with considering systems engineering challenges (Q7) as particularly important and weakly correlated regarding process guidelines (Q16) as a promising solution ($\rho=0.37$). As these are the only correlations with Q4, it indicates that organizations that have reached a certain ML maturity have progressed beyond specific issues and instead focus on the bigger picture, i.e, how to incorporate ML in systems and how to adapt internal processes in the new ML era.

There are more correlations within the categories of challenges (Q5-Q11) and solution proposals (Q12-Q16) than between the two groups. The only strong correlation between groups is test specification (Q11) and formal methods (Q12) ($\rho=0.71$). Within the challenges, the correlation between the two challenges state-space explosion (Q5) and transparency (Q8) stands out as particularly strong ($\rho=0.91$), illustrating the close connection between these two issues with large DNN architectures. Also the two challenge categories requirements specifications (Q9) and test specifications (Q11) are strongly correlated ($\rho=0.71$), in line with a large body of previous work on aligning the two concepts\cite{bjarnason_challenges_2014}.

\subsubsection{Open questions}
The end of the questionnaire contained an open-ended question (Q17), requesting a comment on Fig.~\ref{fig:mapping} and the accompanying findings: ``although few individual V\&V challenges related to machine learning transparency are highlighted in the literature, it is the challenge most often addressed by the previous publications' solution proposals. We also find that the second most addressed challenge in previous work is related to state-space explosion.''

Sixteen out of 49 respondents (32.7\%) provided a free text answer to Q17, representing highly contrasting viewpoints. Eight respondents reported that the findings were not in line with their expectations, whether seven respondents agreed -- one respondent neither agreed nor disagreed. Examples of more important challenges emphasized by the respondents include both other listed challenges, i.e., robustness and requirements specification, and other challenges, e.g., uncertainty of sensor data (in automotive) and the knowledge gap between industry and regulatory bodies (in the process industry). Three respondents answer in general terms that the main challenge of ML-based systems is the intrinsic non-determinism. 

On the other hand, the agreeing respondents motivate that state-space explosion is indeed the most pressing challenge due to the huge input space of the operational environment (both in automotive and railway applications). One automotive researcher stresses that the state-space explosion impedes rigid testing but raises the transparency challenge as well -- a lack thereof greatly limits analyzability, which is a key requirement for safety-critical systems. One automotive developer argues that the bigger state-space of the input domain, the bigger the attack surface becomes -- possibly referring to both adversarial attacks and other antagonistic cyber attacks. Finally, two respondents provide answers that encourage us to continue work along to paths in the SMILE project: 1) a tester in the railway domain explains that the traceability during root cause analyses in ML-applications will be critical, in line with our argumentation at a recent traceability conference\cite{borg_traceability_2017}, and 2) one automotive architect argues that the state-space explosion will not be the main challenge as any autonomous driving will have to be within ``guard rails'', i.e., a solution similar to the safety cage architectures we intend to develop in the next phase of the project.

Seven respondents complemented the survey answers with concluding thoughts in Q18. One experienced manager in the aerospace domain explained: ``What is now called ML was called neural nets (but less sophisticated) 30 years ago.'', a statement that supports our recommendation that the automotive industry should aim for a cross-domain knowledge transfer regarding V\&V of ML-based systems. The manager followed by stating: ``it (ML) introduces a new element in safety engineering. Or at least it moves the emphasis to more resilience. If the classifier is wrong, then it becomes a hazard and the system must be prepared for it.'' We agree with the respondent that actions needed in the hazardous situation must be well-specified. Two respondents comment that conservatism is fundamental in functional safety, one of them elaborates that the ``end of predictability'' introduced by ML is a disruptive change that requires a paradigm shift.

\section{Revisiting the RQs} \label{sec:synth}
This section first discusses the RQs in a larger context, and then aggregates the four sources of evidence presented in Fig.~\ref{fig:evidence}. Finally, we discuss implications for research and practice, including automotive manufacturers and regulatory bodies, and conclude by reporting the main threats to validity. Table~\ref{tab:takeaway} summarizes our findings.

\subsection{RQ1: State-of-the-art in V\&V of safety-critical ML}
There is no doubt that deep learning research currently has incredible momentum. New applications and success stories are reported every month -- and many applications come from the automotive domain. The rapid movement of the field is reflected by the many papers our study has identified on preprint archives, in particular the arXiv.org e-Print archive. It is evident that researchers are eager to claim novelty, and thus struggle to publish results as fast as possible.

While DNNs have enabled amazing breakthroughs, there is much less published work on engineering safety for DNNs. On the other hand, we observe a growing interest as several researchers call for more research on DNN safety, as well as ML safety in general. However, there is no agreement on how to best develop safety-critical DNNs, and several different approaches have been proposed. Contemporary research endeavors often address the opaqueness of DNNs, to support analyzability and interpretability of systems with DNN components. 

Deep learning research is in its infancy, and the tangible pioneering spirit sometimes brings the mind to the Wild West. Anything goes, and there is a potential for great academic recognition for groundbreaking papers. There is certainly more fame in showcasing impressive applications than updating engineering practices and processes.

Safety engineering stands as a stark contrast to the pioneering spirit. On the contrary, safety is permeated by conservatism. When a safety standard is developed, it captures the best available practices to engineer safe systems. This approach inevitably results in standards that lag behind the research front -- safety first! In the automotive domain, ISO~26262 was developed long before DNNs for vehicles was an issue. Without question, DNNs constitute a paradigm shift in how to approach functional safety certification for automotive software, and we do not believe in any quick fixes to patch ISO~26262 for this new era. As recognized by researchers before us, e.g., Salay \textit{et al.}\cite{salay_analysis_2017}, there is a considerable gap between ML and ISO~26262 -- a gap that probably needs to be bridged by new standards rather than incremental updates of previous work. 

Broadening the discussion from DNNs to ML in general, our systematic snowballing of previous research on safety-critical shows a peak of aerospace research between 2002-2007 and automotive research dominating from 2013 and onwards. We notice that the aerospace domain allocated significant resources to research on neural networks for adaptive flight controllers roughly a decade before DNNs became popular in automotive research. We hypothesize that considerable knowledge transfer between the domains is possible now, and plan to proceed such work in the near future.  

The academic literature on challenges in ML-based safety engineering has most frequently addressed state-space explosion and robustness (see Table~\ref{tab:cat} for definitions). On the other hand, the most commonly proposed solutions to overcome challenges of ML-based safety engineering are approaches that belong to formal methods, control theory, or probabilistic methods -- but these appear to be only moderately promising by industry practitioners, who would rather see research on simulated test cases. As discussed in relation to RQ2, academia and industry share a common view on what challenges are important, but the level of agreement on what is the best way forward appears to be less clear.

\subsection{RQ2: Main challenges for safe automotive DNNs}
Industry practice is far from certifying DNNs for use in driverless safety-critical applications on public roads. Both the workshop series and the survey show that industry practitioners across organizations do not know how to tackle the challenge of approaching regulatory bodies and certification agencies with DNN-based systems. Most likely, both automotive manufacturers and safety standards need to largely adapt to fit the new ML paradigm -- the current gap appears not to be bridgeable in the foreseeable future through incremental science alone.

On the other hand, although the current safety standards do not encompass ML yet, several automotive manufacturers are highly active in engineering autonomous vehicles. Tesla has received significant media coverage through pioneering demonstrations and self-confident statements. Volvo Cars is also highly active through the Drive Me initiative, and has announced a long-lasting partnership with Uber toward autonomous taxis. 

Several other partnerships have recently been announced among automotive manufacturers, chipmakers, and ML-intensive companies. For example, Nvidia has partnered with Uber, Volkswagen, and Audi to support engineering self-driving cars using their GPU computing technology for ML development. Nvidia has also partnered with the Internet company Baidu, a company that has a highly competitive ML research group. Similarly, the chipmaker Intel has partnered with Fiat Chrysler Automobiles and the BMW Group to develop autonomy around their Mobileye solution. Moreover, large players such as Google, Apple, Ford, and Bosch are active in the area, as well as startups such as nuTonomy and FiveAI -- no one wants to miss the boat to the lucrative future.

While there are impressive achievements both from spearheading research, and some features are already available on the consumer market, they all have in common that the safety case argumentation relies on a human-in-the-loop. In case there is a critical situation, the human driver is expected to be present and take control over the vehicle. There are joint initiatives to formulate regulations for autonomous vehicles, but, analogously, there is a need for initiatives paving the way for new standards addressing functional safety of systems that rely on ML and DNNs.

We elicited the most pressing issues concerning engineering of DNN-based systems through a workshop series and a survey with practitioners. Many discussions during the workshops were dominated by robustness of DNN components, including detailed considerations about robust DNN architectures and the requirements on training data to learn a robust DNN model. Also the survey shows the importance of ML robustness, which motivates the attention it has received in academic publications (cf. RQ1). On the other hand, while there is an agreement on the importance of ML robustness between academia and industry, how to tackle the phenomenon is still an open question -- and thus a potential avenue for future research. Nonetheless, the problem of training a robust DNN component corresponding to the complexity of public traffic conforms with several of the ``enduring problems'' highlighted by the US Air Force in their technical report on V\&V of autonomous systems~\textbf{[P1]}, e.g., state-space explosion and unpredictable environments.

While robustness is stressed by practitioners, academic publications have instead to a larger extent highlighted challenges related to the limited transparency of ML-based systems (e.g., Bhattacharyya \textit{et al.}\cite{bhattacharyya_certification_2015}) and the inevitable state-space explosion. The survey respondents confirm these challenges as well, but we recommend future studies to meet the expectations from industry regarding robustness research. Note, however, that the concept of robustness might have different interpretations despite having a formal IEEE definition\cite{ieee_computer_society_610.12-1990_1990}. Consequently, we call for an empirical study to capture what industry means by ML and DNN robustness in the automotive context.

The workshop participants perceived two possible approaches to pave the way for safety-critical DNNs as especially promising. First, continuous monitoring of DNN input using a safety cage architecture, a concept that has been proposed for example by Adler \textit{et al.}\cite{adler_safety_2016}. Monitoring safe operation, and re-directing execution to a ``safe track'' without DNN involvement when uncertainties grow too large, is an example of the safety strategy safe fail\cite{varshney_engineering_2016}. Another approach to engineering ML safety, considered promising by the workshops and the survey respondents alike, is to simulate test cases. %Based on these findings, our future research within the SMILE consortium will pursue development of a safety cage architecture encapsulating a DNN component. Moreover, also in line with the view of industry practitioners, we will use state-of-the-art automotive simulators to evaluate our work.

\begin{table*}[]
\centering
\caption{Condensed findings in relation to the research questions, and implications for research and practice.}
\label{tab:takeaway}
\begin{tabular}{|p{5cm}|p{12.5cm}|}
\hline
RQ1. What is the state-of-the-art in V\&V of ML-based safety-critical systems? &
\begin{itemize}
\item Most ML research showcases applications, while development on ML V\&V is lagging behind.
\item Considerable gap between V\&V mandated by safety standards and nature of contemporary ML-based systems.
\item The aerospace domain has collected experiences from V\&V of adaptive flight controllers based on neural networks.
\item Support for V\&V of ML-based systems can be organized into: 1) formal methods, 2) control theory, 3) probabilistic methods, 4) process guidelines, and 5) simulated test cases.
\item Academia has focused mostly on 1)--3), whereas industry perceives 5) as the most promising.
\end{itemize} 
\\
\hline
RQ2. What are the main challenges when engineering safety-critical systems with DNN components in the automotive domain? &
\begin{itemize}
\item How to certify safety-critical systems with DNNs for use on public roads is unclear.
\item Industry stresses robustness, whereas academia most often addresses state-space explosion and the lack of ML transparency.
\item Challenges elicited corroborate work on V\&V by NASA and USAF, covering neural networks, autonomous systems, and adaptive systems.
\end{itemize}
\\
\hline
Implications for research and practice &
\begin{itemize}
\item Gap between ML practice and ISO~26262 requires novel standards rather than incremental updates.
\item Cross-domain knowledge transfer from the aerospace V\&V engineers to the automotive domain appears promising.
\item Need for empirical studies to clarify what robustness means in the context of DNN-based autonomous vehicles. 
\item Systems-based safety approaches encouraged by industry, including safety cage architectures and simulated test cases.  
\end{itemize}
\\
\hline
\end{tabular}
\end{table*}

\section{Conclusion and future work} \label{sec:concl}
Deep learning Neural Networks (DNN) is key to enable the vehicular perception required for autonomous driving. However, the behavior of DNN components cannot be guaranteed by traditional software and system engineering approaches. On top of that, crucial parts of the automotive safety standard ISO~26262 are not well-defined for certifying autonomous systems\cite{salay_analysis_2017,henriksson_automotive_2018} -- certain process requirements contravene the nature of developing Machine Learning (ML)-based systems, especially in relation to Verification and Validation (V\&V). 

Roughly a decade ago, using Artificial Neural Networks (ANN) in flight controllers was an active research topic, and also how to adhere to the strict aerospace safety standards. Now, in the advent of autonomous driving, we recommend the automotive industry to learn from guidelines\cite{pullum_guidance_2007} and lessons learned\cite{taylor_methods_2006} from V\&V of ANN-based components developed to conform with the DO-178B software safety standard for airborne systems. In particular, automotive software developers need to evolve practices for \textit{configuration management} and \textit{architecture specifications} to encompass fundamental DNN design elements. Also, \textit{requirements specifications} and the corresponding \textit{software testing} must be augmented to address the adaptive behavior of DNNs. Finally, the highly \textit{iterative development lifecycle of DNNs} should be aligned with the traditional automotive V-model for systems development. A recent NASA report on safety certification of adaptive aerospace systems\cite{bhattacharyya_certification_2015} confirms the challenges of requirements specification and software testing. Moreover, related to ML, the report adds the \textit{lack of documentation and traceability} in many open source libraries, and the issue of an \textit{expertise gap between regulators and engineers} -- conventional source code in C/C++ is very different from an opaque ML model trained on a massive dataset.

The work most similar to ours also originated in the aerospace domain, i.e., a project initiated by the US Air Force to describe enduring problems (and future possibilities) in relation to safety certification of autonomous systems~\textbf{[P1]}. The project highlighted four primary challenges: 1) state-space explosion, 2) unpredictable environments, 3) emergent behavior, and 4) human-machine communication. While not explicitly discussing ML, the first two findings match the most pressing needs elicited in our work, i.e., \textit{state-space explosion as stressed by the academic literature} (in combination with limited transparency) and \textit{robustness as emphasized by the workshop participants as well as the survey respondents} (referred to as unpredictable environments in~\textbf{[P1]}).

After having reviewed the state-of-the-art and state-of-practice, the SMILE project will now embark on a solution-oriented journey. Based on the workshops, and motivated by the survey respondents, we conclude that \textit{pursuing a solution based on safety cage architectures}\cite{heckemann_safe_2011,adler_safety_2016} encompassing DNN components is a promising direction. Our rationale is three-fold. First, the results from the workshops with automotive experts from industry clearly motivates us, i.e., the participants strongly encouraged us to explore such a solution as the next step. Second, we believe it would be feasible to develop a \textit{safety case} around a safety cage architecture, since the automotive industry already uses the concept in the physical vehicles. Third, we believe the DNN technology is ready to provide what is needed in terms of novelty detection. The safety cage architecture we envision will continuously monitor input data from the operational environment to re-direct execution to a non-ML safe track when uncertainties grow too large. Consequently, we advocate \textit{DNN safety strategies using a systems-based approach} rather than techniques that focus on the internals of DNNs. Finally, also motivated by both the workshops and the survey respondents, we propose an approach to V\&V that makes heavy use of \textit{simulation} -- in line with previous recommendations by other researchers \cite{abdessalem_testing_2016,bojarski_end_2016,tian_deeptest_2018}.

Future work will also study how \textit{transfer learning} could be used to incorporate training data from different contexts or manufacturers, or even include synthetic data from simulators, into DNNs for real-world automotive perception. So far we have mostly limited the discussion to fixed DNN-based systems, i.e., systems trained only prior to deployment. An obvious direction for future work is to to explore how dynamic DNNs would influence our findings, i.e., DNNs that adapt by continued learning either in batches or through online learning. Furthermore, research on V\&V of ML-based systems is more complex than pure technology in isolation. Thus, we recognize the need to explore both ethical and legal aspects involved in safety certification of ML-based systems. Finally, there is a new automotive standard under development that will address autonomous safety: ISO/PAS 21448 Road vehicles -- Safety of the intended functionality. We are not aware of its contents at the time of this writing, but once published, we will use it as an important reference point for our future solution proposals.

% use section* for acknowledgment
\section*{Acknowledgments}
Thanks go to all participants in the SMILE workshops, in particular Carl Zand\'en, Michaël Simoen, and Konstantin Lindstr\"om. This work was carried out within the SMILE and SMILE II projects financed by Vinnova, FFI, Fordonsstrategisk forskning och innovation under the grant numbers: 2016-04255 and 2017-03066.

%\section*{Appendix A}

\begin{table*}[]
\centering
\caption{The start set and the four subsequent iterations of the snowballing literature review.}
\label{tab:snowball}
\begin{tabular}{|p{1.6cm}|p{15cm}|}
\hline
Start set & \textbf{{[}P1{]}} M. Clark \textit{et al.}\cite{clark_air_2014}, \textbf{{[}P2{]}} D. Amodei \textit{et al.}\cite{amodei_concrete_2016}, \textbf{{[}P3{]}} G. Brat and A. Jonsson\cite{brat_challenges_2005}, \textbf{{[}P4{]}} A. Broggi \textit{et al.}\cite{broggi_extensive_2013}, \textbf{{[}P5{]}} B. Taylor \textit{et al.}\cite{taylor_verification_2003}, \textbf{{[}P6{]}} J. Taylor \textit{et al.}\cite{taylor_alignment_2016}, \textbf{{[}P7{]}} A. Carvalho \textit{et al.}\cite{carvalho_automated_2015}, \textbf{{[}P8{]}} S. Ramos \textit{et al.}\cite{ramos_detecting_2016}, \textbf{{[}P9{]}} R. Alexander \textit{et al.}\cite{alexander_situation_2015}, \textbf{{[}P10{]}} X. Zou \textit{et al.}\cite{zou_validation_2016}, \textbf{{[}P11{]}} X. Zou \textit{et al.}\cite{zou_safety_2014}, \textbf{{[}P12{]}} J. Arnold and R. Alexander\cite{arnold_testing_2013}, \textbf{{[}P13{]}} S. Sivaraman and M. Trivedi\cite{sivaraman_active_2014}, \textbf{{[}P14{]}} A. Mozaffari \textit{et al.}\cite{mozaffari_robust_2015} \\
\hline                                                                                                                                                                                                                                                                                                                                                                                                         Iteration 1 & \textbf{{[}P15{]}} S. Seshia \textit{et al.} \cite{seshia_towards_2016}, \textbf{{[}P16{]}} P. Helle \textit{et al.}\cite{helle_testing_2016}, \textbf{{[}P17{]}} L. Li \textit{et al.},\cite{li_intelligence_2016}, \textbf{{[}P18{]}} W. Shi \textit{et al.}\cite{shi_efficient_2016}, \textbf{{[}P19{]}} K. Sullivan \textit{et al.}\cite{sullivan_using_2011}, \textbf{{[}P20{]}} R. Broderick\cite{broderick_adaptive_2005}, \textbf{{[}P21{]}} N. Li \textit{et al.}\cite{li_game-theoretic_2016}, \textbf{{[}P22{]}} S. Russell \textit{et al.}\cite{russell_research_2016}, \textbf{{[}P23{]}} A. Broggi \textit{et al.}\cite{broggi_real_2007}, \textbf{{[}P24{]}} J. Schumann and S. Nelson\cite{schumann_toward_2002}, \textbf{{[}P25{]}} J. Hull \textit{et al.}\cite{hull_verification_2002}, \textbf{{[}P26{]}} L. Pulina and A. Tacchella\cite{pulina_abstraction-refinement_2010}, \textbf{{[}P27{]}} S. Lefevre \textit{et al.}\cite{lefevre_survey_2014}\\
\hline                                                                                                                                                                                                                                                                                                                                                                                                         Iteration 2 & \textbf{{[}P28{]}} X. Huang \textit{et al.}\cite{huang_safety_2016}, \textbf{{[}P29{]}} K. Sullivan \textit{et al.}\cite{sullivan_using_2011}, \textbf{{[}P30{]}} P. Gupta P and J. Schumann\cite{gupta_tool_2004}, \textbf{{[}P31{]}} J. Schumann \textit{et al.}\cite{schumann_toward_2005}, \textbf{{[}P32{]}} R. Broderick\cite{broderick_statistical_2004}, \textbf{{[}P33{]}} Y. Liu \textit{et al.}\cite{liu_validating_2007}, \textbf{{[}P34{]}} S. Yerramalla \textit{et al.}\cite{yerramalla_approach_2005}, \textbf{{[}P35{]}} R. Zakrzewski\cite{zakrzewski_randomized_2004}, \textbf{{[}P36{]}} S. Yerramalla \textit{et al.}\cite{yerramalla_approach_2005}, \textbf{{[}P37{]}} G. Katz \textit{et al.}\cite{katz_reluplex:_2017}, \textbf{{[}P38{]}} A. Akametalu \textit{et al.}\cite{akametalu_reachability-based_2014}, \textbf{{[}P39{]}} S. Seshia \textit{et al.}\cite{seshia_formal_2015}, \textbf{{[}P40{]}} A. Mili \textit{et al.}\cite{mili_towards_2004}, \textbf{{[}P41{]}} Z. Kurd \textit{et al.}\cite{kurd_developing_2007}, \textbf{{[}P42{]}} L. Pulina and A. Tacchella\cite{pulina_never:_2011}, \textbf{{[}P43{]}} J. Schumann \textit{et al.}\cite{schumann_toward_2002}, \textbf{{[}P44{]}} R. Zakrzewski\cite{zakrzewski_verification_2001}, \textbf{{[}P45{]}} D. Mackall \textit{et al.}\cite{mackall_verification_2002}\\
\hline
Iteration 3 & \textbf{{[}P46{]}} S. Jacklin \textit{et al.}\cite{jacklin_development_2005}, \textbf{{[}P47{]}} N. Nguyen and S. Jacklin\cite{nguyen_neural_2007}, \textbf{{[}P48{]}} J. Schumann and Y. Liu\cite{schumann_tools_2007}, \textbf{{[}P49{]}} N. Nguyen and S. Jacklin\cite{nguyen_stability_2010}, \textbf{{[}P50{]}} S. Jacklin \textit{et al.}\cite{jacklin_case_2006}, \textbf{{[}P51{]}} J. Taylor \textit{et al.}\cite{taylor_alignment_2016}, \textbf{{[}P52{]}} G. Li \textit{et al.}\cite{li_scenario-based_2010}, \textbf{{[}P53{]}} P. Gupta \textit{et al.}\cite{gupta_performance_2005}, \textbf{{[}P54{]}} K. Scheibler \textit{et al.}\cite{scheibler_towards_2015}, \textbf{{[}P55{]}} P. Gupta \textit{et al.}\cite{gupta_verification_2004} , \textbf{{[}P56{]}} S. Jacklin \textit{et al.}\cite{jacklin_verification_2004}, \textbf{{[}P57{]}} V. Cortellessa \textit{et al.}\cite{cortellessa_certifying_2000}, \textbf{{[}P58{]}} S. Yerramalla \textit{et al.}\cite{yerramalla_lyapunov_2003}\\
\hline  
Iteration 4 & \textbf{{[}P58{]}} S. Jacklin,\cite{jacklin_closing_2008}, \textbf{{[}P59{]}} F. Soares \textit{et al.}\cite{soares_neural_2006}, \textbf{{[}P60{]}} X. Zhang \textit{et al.}\cite{zhang_controller_2015}, \textbf{{[}P61{]}} F. Soares and J. Burken\cite{soares_flight_2006}, \textbf{{[}P62{]}} C. Torens \textit{et al.}\cite{torens_certification_2014}, \textbf{{[}P63{]}} J. Bosworth and P. Williams-Hayes\cite{bosworth_flight_2007}, \textbf{{[}P64{]}} R. Zakrzewski\cite{zakrzewski_verification_2002}\\
\hline
\end{tabular}
\end{table*}

\section*{References}

\bibliographystyle{plain}
\bibliography{smile}

%\bibliographystyle{plain}
%\begin{thebibliography}{000}
%\bibitem{1}
%J. J. Hopfield, Neurons with graded response have collective computational properties like two-state neurons, {\it Proc. Natl. Acad. Sci.}, {\bf 81}, (1984) 3088--3092.
%
%\bibitem{2}
%D. W. Tank and J. J. Hopfield, Simple `neural' optimization networks: An A/D converter, signal decision circuit, and a linear programming circuit, {\it IEEE Trans. on Circuits and Systems}, {\bf 33}, (1986) 533--541.
%
%\bibitem{3}
%Y. S. Foo and Y. Takefuji, Integer linear programming neural networks for job-shop scheduling, {\it Proc. IEEE Intl. Conf. on Neural Networks}, {\bf II}, (1988) 341--348.

%\bibitem{WS1995}
%B. Widrow and S. D. Steams, \emph{Adaptive Signal Processing}, (Prentice Hall, Englewood, NJ, 1995).
%
%\bibitem{LB1999}
%R. Loren and D. B. Benson (eds.), \emph{Introduction to String Field Theory}, 2\up{nd} edn. (Springer-Verlag, New York, 1999).
%
%\bibitem{K1972}
%R. M. Karp, Reducibility among combinatorial problems, in \emph{Complexity of Computer Computations}, (Plenum, New York, 1972), pp. 85--104.
%
%\bibitem{LB1983}
%R. Loren and D. B. Benson, Deterministic flow-chart interpretations, \emph{J. Comput. System Sci.} \textbf{27} (2) (1983) 400--433.
%
%\bibitem{LLB1983}
%R. Loren, J. Li, and D. B. Benson, Deterministic flowchart interpretations, in \emph{Proc. 3\up{rd} Int. Conf. Entity-Relationship Approach}, eds. C. G. Davis and R. T. Yeh
%(North-Holland, Amsterdam, 1983), pp. 421--439.
%
%%\phantom{00}
%\label{\labart-LastPage}
%\end{thebibliography}
\end{multicols}
\end{document}